%% file: main.tex
\documentclass[11pt]{article}
\usepackage{graphicx}
\usepackage{url}
\usepackage{mathtools, amssymb, amsthm, enumerate}
\usepackage{algorithmic}
\usepackage{textcomp}
\usepackage{adjustbox}
\usepackage[table,xcdraw]{xcolor}
\usepackage[utf8]{inputenc}
\usepackage[mathletters]{ucs}
\usepackage{mathptmx}
\usepackage{anyfontsize}
\usepackage{caption}
\usepackage{subcaption}
\usepackage{float}
\usepackage{amssymb}
\usepackage{pifont}
\usepackage{fancyhdr}
\usepackage{lastpage}

\definecolor{linkcolor}{rgb}{0.2, 0.2, 0.6}
\definecolor{referencecolor}{rgb}{0.0, 0.5, 0.0}
\definecolor{highlight}{rgb}{0.96, 0.96, 0.86}

\def\sysname{\textsc{PerOS}}

\newcommand{\tocite}[1]{[{\textcolor{red}{?}}]}
\newcommand{\toref}[1]{\textcolor{red}{N}}

\usepackage{tikz}
\newcommand{\circleb}[1]{%
 \begin{tikzpicture}[baseline=(char.base)]
   \node[draw,circle,inner sep=0.5pt, fill=black, text=white] (char){\small {#1}};
 \end{tikzpicture}%
 }
 \newcommand{\circlew}[1]{%
 \begin{tikzpicture}[baseline=(char.base)]
   \node[draw,circle,inner sep=0.5pt, fill=white, text=black] (char){\small {#1}};
 \end{tikzpicture}%
 }

 \newcommand{\circley}[1]{%
 \begin{tikzpicture}[baseline=(char.base)]
   \node[draw,circle,inner sep=0.5pt, fill=yellow, text=black] (char){\small {#1}};
 \end{tikzpicture}%
 }
\newcommand{\mypar}[1]{{\noindent\bf #1.\ }}

\usepackage[
  urlcolor=linkcolor,
  citecolor=referencecolor,
  breaklinks,
  unicode
]{hyperref}

\theoremstyle{remark}

\usepackage[numbers]{natbib}

\usepackage{listings}
\lstnewenvironment{mylisting}[1][]%
{\lstset{frame=tb, framerule=1pt, captionpos=b, caption=#1}}%
{}
\usepackage{enumitem} 

\begin{document}

\thispagestyle{empty}

\begin{center}


{\Large Research Vision}

\vspace*{3.5cm}

\rule{1\linewidth}{5pt}\\[0.4cm]
{\fontsize{25}{16} \bfseries \sysname: Personalized Self-Adapting \par} 
\medskip
{\fontsize{25}{16} \bfseries Operating Systems in the Cloud \par}\vspace{0.4cm}
\rule{1\linewidth}{1.5pt}\\[1.5cm]



\vspace*{2.5cm}

{\large
    \begin{tabular}{c}
    \\ 
    \Large\textbf{Hongyu H\`e} \\
    
    {\textit{ETH Z\"urich}} \\
    \texttt{hongyu.he@inf.ethz.ch}
    \end{tabular}
}


\vspace*{5cm}



November 2023\\[4cm] 

\end{center}

\newpage

\begin{abstract}

Operating systems (OSes) are foundational to computer systems, managing hardware resources and ensuring secure environments for diverse applications. However, despite their enduring importance, the fundamental design objectives of OSes have seen minimal evolution over decades. Traditionally prioritizing aspects like speed, memory efficiency, security, and scalability, these objectives often overlook the crucial aspect of intelligence as well as personalized user experience. The lack of intelligence becomes increasingly critical amid technological revolutions, such as the remarkable advancements in machine learning (ML).

Today's personal devices, evolving into intimate companions for users, pose unique challenges for traditional OSes like Linux and iOS, especially with the emergence of specialized hardware featuring heterogeneous components. Furthermore, the rise of large language models (LLMs) in ML has introduced transformative capabilities, reshaping user interactions and software development paradigms.

While existing literature predominantly focuses on leveraging ML methods for system optimization or accelerating ML workloads, there is a significant gap in addressing personalized user experiences at the OS level. To tackle this challenge, this work proposes \sysname{}, a personalized OS ingrained with LLM capabilities. \sysname{} aims to provide tailored user experiences while safeguarding privacy and personal data through declarative interfaces, self-adaptive kernels, and secure data management in a scalable cloud-centric architecture; therein lies the main research question of this work: How can we develop intelligent, secure, and scalable OSes that deliver personalized experiences to thousands of users?
\end{abstract}

\newpage

\tableofcontents

\newpage

\input{sections/1_intro}

\input{sections/2_interface}

\input{sections/3_kernel}

\input{sections/4_deploy}

\input{sections/5_related_work}






\newpage

\cleardoublepage

\phantomsection

\addcontentsline{toc}{section}{References}

\bibliographystyle{abbrvnat}

\bibliography{main}

\end{document}

%% file: sections/1_intro.tex
\section{Introduction} \label{sec:intro}

The complexity and significance of operating systems (OSes) cannot be overstated. 
They serve as the backbone in nearly all computer systems, effectively managing hardware resources and ensuring a secure environment for running various applications. 
Despite their pivotal role, the fundamental design objectives of OSes have seen minimal evolution over almost half a century. 
These objectives traditionally prioritize aspects like speed, memory efficiency, security, scalability, consistency, concurrency control, and fault tolerance in distributed settings.
However, this approach predominantly views OSes as general-purpose computing infrastructure, often overlooking the individual user experience. 
This oversight poses a risk, particularly in the face of the ongoing technological revolution spurred by remarkable advancements in machine learning (ML).

Firstly, personal devices like mobile phones and laptops are forging increasingly intimate connections with their users. 
These gadgets transcend mere computing tools; they are companions, or even perceived as friends or even family members, safeguarding precious memories and engaging in daily interactions. 
In today's landscape of diverse applications, individual users harbor unique needs, each representing specific workloads. 
For instance, some devote substantial time to immersive video gaming or AR/VR applications, demanding substantial computational resources, while others predominantly engage in web browsing and text editing.
Moreover, to cater the ever-growing resource demands in the post-Moore's Law era, hardware is becoming increasingly specialized, having heterogeneous components integrated onto a single chip (e.g., AMD Versal~\cite{versal} and Apple M3~\cite{m3}). 
Yet, adapting mature OSes such as Linux and IOS to these fast evolving devices poses challenges, partly due to their monolithic design, wherein subsystems closely coupled for better performance.
These converging trends have been expedited by the rise of ML, notably, large language models (LLMs).

While training and tuning several LLMs at Apple each on hundreds of TPUs, I noticed their fundamental differences from the traditional ML models ---
they absorb web-scale data and can pick up a data point after only a few exposures; 
they feature \textit{in-context learning}, customizing their responses based on prior interactions; 
they are capable of \textit{few-shot learning}, carrying out similar tasks given just a few examples; 
they can \textit{follow instructions}, tailoring their output given users' prompts. 
Furthermore, LLMs have been extended to \textit{multimodal input}, such as images and video, performing multiple desperate tasks with end-to-end trained architectures. 
With their exceptional language understanding and data manipulation capabilities, LLMs effortlessly process files of various formats and comprehend how to manipulate the data that documents contain. 
These capabilities have drastically changed how users interact with their applications, as much as the ways in which software is built. 
I believe that if the iPhone was the great hardware filter, then the LLM is going to be a great filter for software, including OSes.

Unfortunately, the literature has been primarily focused on either leveraging ML methods for automated system optimization (e.g., \cite{Pan2019LateBR,Alawieh2020HighDefinitionRC,Reza2018NeuroNoCEO,Boyan1993PacketRI,Joardar2018LearningBasedA3,Zhang2018LearningDP,Zhang2021SinanMA,Zhou2022AQUATOPEQR,Wang2022SOLSO,kruit2020tab2know}) or specializing systems for accelerating ML workloads (e.g, \cite{moritz2018ray,Tian2017ELFAE,Gauci2018HorizonFO,Kaler2023CommunicationEfficientGN,Kharbanda2022CascadeXMLRT,Wang2023FLINTAP}).
This inclination neglects the importance of individual users embodying distinct requirements and usage patterns that require personalized adaptations from OSes.
Similarly, while most technology companies I have worked with (e.g., Huawei, Microsoft, and Oracle) have been employing ML models to ease the burden of monitoring and/or orchestrating large-scale resources, `personalized intelligence’ on the OS level has not yet gained prominence (apart from a few general solutions like Cortana and Siri).
These aforementioned challenges give rise to the main research question (\ref{mrq}) of this proposal:

\begin{enumerate} [label=\textbf{MRQ}]
 \item \label{mrq} How to build intelligent, secure, and scalable OSes that provide thousands of users with smooth and personalized experiences?
\enlargethispage{\baselineskip}
\end{enumerate}

As a first step towards attacking the \ref{mrq}, I propose \sysname{},\footnote{\sysname{} is just a working code name.} a personalized LLM-ingrained OS capable of accommodating the needs and usage patterns of individual users while protecting their privacy and personal data.
\sysname{} aims to offer a declarative user interface~(\S\ref{sec:interface}), self-adaptive kernel~(\S\ref{sec:kernel}), and secure management of personal data with a scalable cloud-centric architecture~(\S\ref{sec:deploy}).


%% file: sections/2_interface.tex
\section{Declarative User Interface} \label{sec:interface}


Throughout history, human-computer interaction has evolved in distinct phases. 
Initially, users grappled with systems like punch cards and command-line interfaces (CLI), forced to adapt to computers' rigid expectations. 
The emergence of Graphical User Interfaces (GUIs) marked a pivotal shift, making computers more accessible to non-technical users.
However, GUIs, while transformative, only partially bridged the human-computer gap. 
Users still had to search among hundreds of applications and navigate preset layouts that mirror developers' logic, leading to frustrating experiences of hunting for specific functionalities. 
This reliance on thinking like developers curtailed the user experience, imposing rigid behavior patterns.
That been said, such unpleasant experiences also apply to developers themselves.
For example, CLI and writing scripts remain the primary means for compiling and testing applications, which are typically error-prone and hard to maintain (as seen in composing Shell scripts and chaining commands).
Fortunately, the advent of LLMs signifies a groundbreaking leap --- they enabled conversational interfaces that \textit{require no behavioral shifts from users}, enabling intuitive and smooth user experience.

Integrating LLMs into existing systems automates functionalities, enriching capabilities and simplifying interactions. 
An exemplar is Microsoft's Windows Copilot~\cite{copilot}, a centralized AI assistant empowering users within the Windows environment. 
Seamlessly accessible from the taskbar, Copilot acts as a personal assistant, enabling content rewriting and summarization alongside traditional Windows features. 
Similarly, the recently unveiled GPT4-turbo~\cite{gpt4} showcases expanded functionalities like sending Slack messages, setting calendar events, and invoking various software APIs, which highlights the immense potential of LLMs as a unified user interface.
Such advancements can radically blur the line between applications and the OS.



\subsection{Objectives} \label{subsec:obj1}
In \sysname{}, I want to build a declarative, unified interface powered by LLMs to facilitate the interaction between the user and the OS.
Specifically, users express the high-level tasks that they want to perform in natural languages like English.
Then, \sysname{} performs the tasks and returns the results to the users.
Listing~\ref{list:request} shows an example user request. 
In this example, \sysname{} needs to extract the tasks from the user request in the correct order, navigate to the intended project directory, carry out the tasks with the corresponding API calls, double-check the result with the user, and ask clarification questions in case of any ambiguity.
Ideally, this interface could be extended to multi-modalities, such as speech and images.
Such an extension would foster improved accessibility by allowing users to interact with the system through their preferred modalities, catering to various communication preferences.
Although this naive example can be achieved via a chain of commands in a rule-based system, hand-crafted systems typically require users to structure their queries in rigid formats in order to correctly extract information such as target objects and destination.
Moreover, developers face challenges in anticipating all request patterns, leading to an ever-expanding set of rules and templates in the knowledge base to accommodate new types of user requests.
Likewise, the evolution of system interfaces and application APIs brings about changes over time. Introducing or modifying utilities might necessitate updating all associated functionalities that rely on them. This constant evolution demands ongoing adjustments to maintain synchronization across the system.
The benefit of employing LLMs, therefore, lies in their ability to ensure a seamless user experience and enable continuous learning. They excel in adapting to new system APIs and unforeseen user requests by generalizing from just a few examples. This capability streamlines the adaptation process and promotes agility in accommodating changes within the system landscape.

\input{assets/list_request}

\subsection{Research Question and Challenges} 
The objectives described above raise the first RQ of this proposal:
\begin{enumerate} [label=\textit{RQ1}]
 \item \label{rq1} How can we enable users to interact with their OSes using natural language in a declarative fashion, fostering personalized experiences and seamless interaction?
\end{enumerate}

\mypar{Challenges}
I anticipate the following challenges in answering \ref{rq1}.
Firstly, collecting high-quality training data is essential to ensuring model quality. 
Specifically, training requires thousands of request-solution pairs.
Next, based on my experience in working with LLMs, achieving concise answers is a non-trivial task, as existing LLMs are not guaranteed to be self-consistent~\cite{Wang2022SelfConsistencyIC}.
In essence, LLMs do not consistently produce the end-of-sequence token and often rely on predefined stopping probability thresholds.
Consequently, fine-tuning is necessary to obtain brief answers, as illustrated in Listing~\ref{list:request}.
Similarly, maintaining a continuous and coherent dialogue over time poses a challenge, particularly during multi-stage tasks (akin to the scenario in Listing~\ref{list:request}).
Additionally, integrating the developed LLM with a kernel (e.g., Linux) requires substantial engineering effort. 
For instance, efficient interactions with the LLM may necessitate a kernel module, since the model is expected to run in user mode.
Moreover, registering event triggers, monitoring kernel updates and communicating the relevant updates to the LLM \textit{in a format consistent with the training stage} are crucial.
Also, determining the optimal timing and strategy for model retraining poses significant challenges.
The last challenge lies in defining representative metrics to evaluate the LLM-integrated system. Evaluating LLMs remains an ongoing research area, making the selection of suitable metrics tricky.
Common automatic metrics like BLEU, ROUGE, and METEOR might not capture this system's performance well. Here, not only linguistic quality but also the relevance of the solution and the accuracy of its execution are paramount. These factors pose critical evaluation criteria that automatic metrics might not wholly encompass.

\subsection{Research Methods} \label{subsec:method1}

\mypar{Scope}
A full-fledged system as described in \S\ref{subsec:obj1} is appealing but overly extensive in scope. To make addressing \ref{rq1} feasible within a year, I narrow it down in the following ways.
Firstly, restricting the user interface solely to text format is essential, considering that multimodality is currently the research frontier of ML modeling --- a domain not central to this systems research. 
The proof-of-concept for \ref{rq1} would already serve as a natural passage to exploring other modalities, such as images and audio.
Secondly, I propose a controlled expansion of the number of applications and APIs. 
Starting with traditional POSIX syscalls and Shell scripting could establish as a solid foundation, allowing for the gradual addition of a few application APIs at a later stage.

\smallskip
\mypar{Approach}
To address \ref{rq1}, I will conduct \textit{development and application studies} to create a prototype system. 
This prototype will serve as the initial incremental step toward building \sysname{}. 
Figure~\ref{fig:interface} provides an overview of the proposed system prototype.
The process begins with the collection and potential synthesis of sufficient training data. 
This dataset should include example user requests, corresponding APIs to be invoked, and the expected system updates returned from the kernel.
Upon acquiring the datasets, the next phase involves training and fine-tuning the LLM. 
I plan to utilize a two-tower LLM, where the encoder functions as the Interpreter (\circleb{1}) processing user requests, and the decoder acts as the Director (\circleb{2}) generating a sequence of operations in response. 
These components should undergo separate evaluations (\S\ref{subsec:eval1}) before integration into the broader system.

In parallel, the Actuator (\circleb{3}) can be constructed to execute the sequence of operations generated by the Director. 
It will invoke the APIs and execute commands by engaging the kernel, specifically employing an off-the-shelf monolithic kernel like Linux.
For \ref{rq1}, I plan to use the kernel only as a static toolbox and do not modify its functionality other than registering event triggers and implementing the kernel module.
The Watchdog (\circleb{4}) assumes the responsibility for monitoring registered kernel events and relaying updates to the Actuator, which, in turn, organizes and forwards results back to the user terminal.
I plan to orchestrate these components as microservices using frameworks familiar to me (e.g., Kubernetes~\cite{k8s} or Service Weaver~\cite{weaver}), employing remote procedure calls (RPC) for efficient inter-service communication. 
Furthermore, \textit{low-end} networking-enabled devices should adequately support the declarative interface for users to send requests to \sysname{}, given that the bulk of computation occurs remotely.
Finally, the LM Manager (\circleb{5}) will evaluate and retrain the LLM according to predefined retraining policies.
 
\input{assets/fig_interface}

\subsection{Evaluation} \label{subsec:eval1}
To comprehensively evaluate the prototype during development, I intend to conduct both \textit{quantitative evaluations} and a \textit{hybrid user study}. 
This study is, to my knowledge, the first aiming to leverage LLMs for an OS user interface. 
Hence, potential baselines could include a separately constructed rule-based system (e.g., \cite{Etzioni1993agentos,Etzioni1994softbot}) or integrating the prototype with traditional sequence models (e.g., Bidirectional LSTMs).

The evaluation should be structured into steps to facilitate component integration and ablation studies. 
Initially, for the isolated evaluation of the LLM, I will employ traditional linguistic metrics (e.g., BLEU, ROUGE, and METEOR) to assess both the Interpreter and Director's language understanding capabilities. 
This step ensures correct information extraction and smooth response generation. 
Subsequently, assessing the Director via accuracy (i.e., correctness of generated commands) and recall (i.e., retrieval rate of necessary commands) is critical. 

Before integrating the Director with the Actuator, the rest of the system should undergo similar testing, including unit testing for command execution, event triggering, and reporting. 
After the integration, a comprehensive end-to-end evaluation is necessary, repeating linguistic and execution evaluations, in which the Actuator takes real Director outputs, and the Director responds based on the actual results from the Watchdog.

The user study aims to gauge efficiency (e.g., speed) and satisfaction levels. 
Student volunteers or commercial platforms like Amazon Mechanical Turk~\cite{turk} may be utilized for this purpose, where evaluators are tasked with performing predefined user requests through three methods: \textit{(i)} using the \sysname{} prototype, \textit{(ii)} a baseline system, and \textit{(ii)} manual execution (with internet access). 
Time taken and qualitative feedback for each category will be recorded and compared.

Lastly, I plan to evaluate the retraining policies used by the LM Manager, by gradually extending the functionality of the prototype.
Specifically, the linguistic quality and the command execution scores are continuously measured as new system APIs and event triggers are added over time. 
System performance initially may deteriorate upon the addition of new utilities and gradually recovers by triggering retraining and learning the added new APIs.

%% file: assets/list_request.tex
\begin{mylisting}[Example user interactions with \sysname{} via the declarative interface., escapeinside={<@}{@>}, float, floatplacement=t, label=list:request]
<@\textbf{User}: now, undo the most resent commit for my HappyDog project, remove all the CSV files larger than 10 MB from the git cache, move those files to a new directory called data at the project root, ignore this folder in git, add a suffix \_large to all their names, augment the previous commit without a new message, and force push to my remote repo@>

<@\textbf{\sysname{}}: The following is the git diff after the changes:@>
index 6b0c6cf..b37e70a 100644
--- a/dogs.csv
+++ b/data/dogs_large.csv
...
<@\textbf{\sysname{}}: Is the result correct? Otherwise, I can revert the changes.@>

<@\textbf{User}: y @>

<@\textbf{\sysname{}}: To which branch do you want to push the changes?@>

<@\textbf{User}: i forgot the remote branches @>

<@\textbf{\sysname{}}: You have \texttt{main}, \texttt{dev}, and \texttt{feat/chihuahua} in your GitHub repo and only \texttt{master} on Bitbucket. @>

<@\textbf{User}: dev github @>

<@\textbf{\sysname{}}: I've pushed the changes to \texttt{https://github.com/username/happydog/tree/dev} @>
\end{mylisting}

%% file: assets/fig_interface.tex
\begin{figure}[t]
    \centering
    \begin{adjustbox}{width=0.91\linewidth,center=0pt}
      \includegraphics[width=\linewidth]{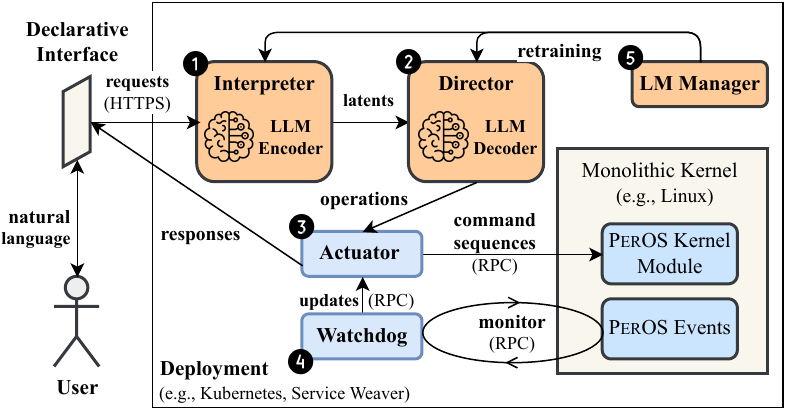}
    \end{adjustbox}
    \caption{Prototype of \sysname{} with the declarative interface powered by LLMs.}
    \label{fig:interface}
\end{figure}

%% file: sections/3_kernel.tex
\section{Adaptive Kernel and its Subsystems} \label{sec:kernel}

The surge in resource demands from emerging software applications, coupled with the increasing heterogeneity of hardware, renders the design and development of OSes without integrating `intelligence' unsustainable. 
Consider Linux, already comprising over 28 million lines of code, which grows bulkier as additional features are incorporated to meet evolving requirements within its general-purpose design. 
Tuning and debugging such a colossal system have become akin to an arcane art, given the intricate interconnections among system components.
For instance, Linux boasts more than 17K kernel configurations, complicating its adaptation to new applications, such as AI workloads, and evolving user demands. 
Consequently, developers have resorted to rewriting, patching, or even bypassing this complexity to directly access the required functionalities.

Traditionally, after building, installing, and configuring an OS, it remains largely static at runtime, limiting its adaptability to the dynamic operating environment. 
To match the swift evolution of modern hardware and applications, I advocate for a different approach in future OS development. 
Rather than relying solely on human expertise, heuristics, and huge amounts of engineering effort, leveraging ML techniques to automatically ``learn'' OS configuration and tuning becomes promising. 
Given the super-human performance we have witnessed in the recent ML development, such learned solutions could potentially yield superior results.
My proposal in this section entails employing ML models as a shim layer that helps tune and configure the underlying kernel to personalize it toward specific user workloads.

\subsection{Objectives} \label{subsec:obj2}

Unlike \S\ref{sec:interface} where a general-purpose kernel was used as a static toolbox, now I want to make the kernel and its subsystems in \sysname{} ``dynamic,'' i.e., they configure and tune themselves to adapt to the users' usage patterns.
I identify three primary areas where ML solutions hold significant promise, each posing an increased level of difficulty.

The first area involves \textit{adaptive configuration and tuning}. 
Present-day kernels like Linux offer numerous tunable knobs. 
For instance, the memory subsystem, filesystem, and networking subsystem individually possess 89, 351, and 729 configurations, respectively. 
These configurations mostly fall into two categories: timing parameters and sizes. 
Timing parameters involve elements like the time slice of kernel tasks, memory swapping frequency in paging, and CPU frequency sampling rates (e.g., for DVFS). 
Likewise, various sizes, such as buffer cache size, disk/swap prefetching data amounts, and memory page sizes with \texttt{hugetlbfs}, can be configured in Linux. 
Many of these configurations significantly impact the performance and energy costs of user applications. 
For example, increased CPU interruption offers potential for improved CPU utilization through more aggressive thread scheduling but may lead to performance overhead by preempting and context switching threads. 
Similarly, a larger buffer cache enhances storage system performance but reduces the amount of available memory for user applications.
Rather than relying solely on labor-intensive engineering efforts, ML models trained on historical user traces could generate configurations adaptive to user usage patterns.

The second area revolves around \textit{learning system policies}, to name a few, memory allocation policies determine which free space is allocated to applications; 
CPU scheduling algorithms dictate which kernel tasks to prioritize; 
cache replacement policies decide which data to evict. 
While these policies have a more direct impact on application performance compared to configurations, they still primarily rely on simple algorithms and heuristics. 
For instance, the \texttt{ext} family of filesystems in Linux allocates adjacent space for files in the same directory, which is only effective for accessing nearby files. 
The most prevalent cache replacement policy is quasi-LRU and aims to swap out the least recently used data.
Apparently, its efficiency heavily relies on data locality. 
Similarly, the Linux CFS (Completely Fair Scheduling) algorithm, utilizing a red-black tree to manage timeshare of tasks, not only incurs a $O(\log N)$ time complexity but also falters under short-task threshing scenarios, where approximations of shortest-remaining-processing-time (SRPT)~\cite{Bansal2001AnalysisOS} could improve latency~\cite{Isstaif2023TowardsLL}. 
My intent is to leverage ML methods to dynamically adjust these policies or even learn new policies tailored to user workloads.

While the first two categories are performance-influential, they typically do not jeopardize kernel execution correctness.
The third area, however, is the most perilous: \textit{learning system functionalities}. 
Inspired by the work on learning database index~\cite{Kraska2017TheCF}, I speculate that similar strategies could be applied to many similar kernel functionalities. 
For example, memory translation, involving mapping virtual memory addresses to physical addresses by traversing multi-level page tables, constitutes a performance bottleneck in virtualization techniques~\cite{Stojkovic2022ParallelVM}, where extended page tables, up to four levels deep, result in severe TLB miss penalties. 
Similarly, indexing into a file's data block requires traversing a multi-level index structure (e.g., the \texttt{ext4} filesystem). 
These functionalities epitomize areas where ML methods could yield improved time and space solutions. 
However, employing ML in this category is exceedingly risky, as erroneous kernel functionalities could crash the system. 
Therefore, instead of replacing existing kernel functionalities, ML models could serve as a first-order approximation, reducing the search space and transforming a global indexing problem into a localized one.

\subsection{Research Question and Challenges} 
The aforementioned objectives give rise to the second RQ of this proposal:
\begin{enumerate} [label=\textit{RQ2}]
 \item \label{rq2} How to make the OS kernel and its subsystems self-adaptive to users' usage patterns by automatically learning from user activities over time?
\end{enumerate}

\mypar{Challenges}
As hinted in \S\ref{subsec:obj2}, addressing \ref{rq2} may encounter several roadblocks.
Primarily, unlike \ref{rq1}, incorporating ML methods closer to the kernel necessitates crucial considerations of performance and space overhead. 
This requirement makes leveraging LLMs unlikely for \ref{rq2} due to their slow inference time (a few milliseconds per token) and substantial memory requirements. 
Even traditional ML methods face challenges due to the stringent latency demands for kernel-level decisions (single-digit microseconds). 
For instance, invoking a model on the present-day PCIe setups usually takes tens of microseconds, making this a potential bottleneck.
Therefore, the ML models employed nearby kernel components has to be \textit{lean and amenable to hardware acceleration}. 

Moreover, even if smaller ML models can fit into the kernel, their execution in kernel mode is improbable due to the lack of ML runtime support and the complexity of kernel integration. 
Additionally, akin to multimodality, multitasking poses an equal challenge in my experience, therefore, employing a single global model for various configurations and policies may be infeasible. 
Hence, determining the appropriate model type for diverse tasks is pivotal.
Furthermore, regardless of the type of ML methods employed, their probabilistic nature conflicts with the high precision demanded by OSes. 

Lastly, evaluating the learning process is the hardest for three primary reasons: 
\textit{(i)} Workload-dependent ground truth complicates many learning tasks, such as determining optimal cache sizes and swapping frequencies.
For \ref{rq2} specifically, another challenge is obtaining several workload traces of consistent usage patterns, mimicking different users. 
\textit{(ii)} Correlated factors add complexity; for example, buffer cache size, flushing frequency, and cache policy collectively impact buffer cache performance, subsequently affecting both the storage and memory subsystems.
\textit{(iii)} System optimization requires a holistic approach; addressing individual components in silos is similar to trying to cure a cold by only treating the symptom of coughing.

\subsection{Research Methods}

\input{assets/list_filesys}

\mypar{Scope}
As outlined in \S\ref{subsec:obj2}, the scope of exploration for \ref{rq2} is extensive.
Diverging from prior work (\S\ref{sec:related}), my focus lies in integrating intelligence into the system as a whole, concentrating on adaptiveness to users' usage patterns, rather than honing in solely on enhancing the performance of individual components such as prefetching~\cite{Shi2019LearningET} and branch prediction~\cite{Calder1997EvidencebasedSB,Jimnez2001DynamicBP,Amant2008LowpowerHA}.
Therefore, the learning tasks should collectively serve as a proof-of-concept, showcasing \sysname{}'s capacity to dynamically adapt to user workloads and fine-tune the kernel setup, improving the utilization of system resources.
Ideally, these tasks should expand across the OS stack, encompassing all three categories described above (\S\ref{subsec:obj2}).
Since the utilization of ML methods in power management, scheduling, and the memory subsystem constitutes a relatively crowded research area (\S\ref{sec:related}), my intention is to concentrate on adding adaptability into the storage and filesystem through learning methods.
This choice of focus also reflects the fact that the filesystem is one of the most intimate components of the OS, with which users interact frequently.
For example, it should seamlessly integrate with the declarative interface introduced in \S\ref{sec:interface}, helping users maintain and manage their files (Listing~\ref{list:filesys}).

\input{assets/fig_kernel}

\smallskip
\mypar{Approach} \label{subsec:method2}
Similar to addressing \S\ref{rq1}, I will be conducting \textit{development and application studies}, extending the initial \sysname{} prototype (\S\ref{subsec:method1}).
However, delving into \ref{rq2} necessitates a closer examination of the kernel (Figure~\ref{fig:kernel}), which was initially utilized as an off-the-shelf, monolithic toolbox. 
Here, I intend to adopt a microkernel such as MINIX~\cite{tanenbaum2010minix}, DBOS~\cite{Skiadopoulos2021DBOSAD}, or seL4~\cite{heiser2020sel4}. 
These microkernels effectively decouple and deprivilege the filesystem from the kernel. 
This step is necessary as it mitigates inherent risks associated with learning system functionalities (\S\ref{subsec:obj2}) and circumvents the need for hosting any ML models in kernel space.

For learning configurations and automatic tuning, I want to use ML methods to dynamically set the disk prefetching amount, a.k.a., the read-ahead buffer size on Linux via \texttt{blockdev} for example.
Large read-ahead buffer sizes can enhance read performance for sequential disk accesses, reducing I/O wait times by prefetching data into memory before it is requested.
User applications performing sequential reads, such as video streaming or large file transfers, benefit the most from increased read-ahead buffer sizes.
However, for workloads involving random or non-sequential data access, large read-ahead buffer sizes might not improve performance --- prefetching large amounts of data ahead of requests is less effective when the access pattern is irregular.
Another trade-off is memory --- while beneficial for sequential reads, allocating excessive memory to prefetching might lead to unnecessary memory overhead, affecting overall system performance.

For investigating learning system policies, I aim to explore a correlated task: learning filesystem space allocation policies. 
Ideally, this learning task should synergize with the learned configuration of the read-ahead buffer size.
Filesystem space allocation policies, such as block allocation strategies (e.g., extent-based allocation or block group allocation), primarily dictate how data is organized and stored on disk. 
ML models can contribute to more intelligent and efficient data placement.
For example, by learning user access patterns, the model can allocate free space for faster file access, such as grouping related files together or reorganizing directories.
Another example is dynamic file tiering, where files are classified based on usage patterns and are automatically moved to different storage tiers for better performance and storage utilization.

Certain filesystems provide mount options or features that exert influence over space allocation strategies or behavior. 
For instance, \texttt{ext4} offers options like \texttt{mballoc} or \texttt{inode\_readahead\_blks} that can impact allocation policies or the read-ahead behavior of inodes.
This relationship is reciprocal --- setting the read-ahead buffer size can also be base on filesystem space allocation policies. 
Considering how different buffer sizes align with the filesystem's characteristics and the user's usage patterns becomes essential.

For investigating system functionality, I plan to study the feasibility of learning multi-level indexing (e.g., inode structures) for mapping file names to disk block addresses.
In Linux \texttt{ext4}, each file is linked to an inode that contains metadata (e.g., permissions, timestamps, pointers to data blocks). 
The filesystem manages these inodes in an inode table, a data structure housing these file metadata entries. 
Each inode retains crucial information about the location and structure of file data blocks.
Within the inode, the filesystem utilizes a hierarchy of block pointers to facilitate the mapping from file names/offsets to disk block addresses.
These pointers can be direct or indirect (double or triple) pointers pointing to data blocks. 
When seeking a specific offset within a file, the filesystem computes the corresponding logical block address by leveraging the inode's block pointers alongside the file's offset.
To navigate through this multi-level index structure, the filesystem relies on those direct and indirect block pointers, aiming to index the relevant data block holding the requested file content.
Learning this intricate mapping may be challenging due to its inherent complexity and the absence of easily identifiable patterns.
Hence, my initial approach would involve generating a dataset that encompasses file names, offsets, and their corresponding disk block addresses. 
This dataset could be created by systematically traversing filesystem structures.
Subsequently, I plan to extract meaningful features from file names, offsets, and potentially file metadata (e.g., file sizes, access patterns, inode details). 
These extracted features will form the base for representing the input data in this complex mapping modeling.

\subsection{Evaluation} \label{subsec:eval2}

The evaluation methodology for the proposed approach (\S\ref{subsec:method2}) predominantly comprises \textit{quantitative studies} involving multiple micro-benchmarks. 
To my knowledge, this study would mark the first exploration into intelligent OS storage and filesystems across all three categories (\S\ref{subsec:obj2}). 
Therefore, a reasonable (and perhaps a strong enough) baseline could be a configuration without ML-based adjustments. 
To conduct this evaluation, I intend to first collect a selection of pertinent benchmarks (e.g., Filebench~\cite{tarasov2018filebench}) and datasets pertinent to storage and filesystems (e.g., UMass I/O Trace~\cite{weibel2013umasstracerepository}), including tools like FIO~\cite{iotester}.
The evaluation process encompasses an array of performance metrics such as throughput, latency, and I/O efficiency, covering diverse workloads comprising read-heavy, write-heavy, sequential, and random access patterns.
For instance, one aspect I aim to evaluate involves measuring the influence of read-ahead buffer size adjustments on system memory usage, ensuring that memory consumption remains within acceptable thresholds. 
Similarly, I plan to assess the impact of changes in filesystem space allocation policies on disk space usage, fragmentation, and overall storage efficiency.
In addition to performance analysis, evaluating the system's stability under varied workloads and scenarios is critical to ensuring that ML-driven adjustments uphold system stability and do not precipitate unexpected failures. 
Stress-testing forms an integral part of this evaluation as well, in which I will examine \sysname{}'s capacity to manage increased workloads, ensuring that ML-based adjustments can seamlessly adapt to shifting demands.
Lastly, similar to \S\ref{subsec:method1}, an important aspect of the evaluation process involves assessing \sysname{}'s responsiveness to alterations in user workload patterns, and its efficiency in dynamically adapting ML-based adjustments accordingly.

%% file: assets/list_filesys.tex
\begin{mylisting}[Example of personalized recommendation through the declarative interface  (\S\ref{sec:interface}) based on recent activities in the filesystem., escapeinside={<@}{@>}, float, floatplacement=t, label=list:filesys]
<@\textbf{\sysname{}}: You have recently created many files in the \texttt{/Documents/crucial/} folder. Should I add them to the backup workflow?@>

<@\textbf{User}: sg! they're important personal data @>

<@\textbf{\sysname{}}: Great, I've added them to the backup list and labeled them as \texttt{important}. They're scheduled for weekly backups, aligning with your workflow's highest frequency. Also, for privacy reason, they'll be stored on your local SSD instead of the cloud. Any other changes you'd like?@>

...
\end{mylisting}

%% file: assets/fig_kernel.tex
\begin{figure}[t]
    \centering
    \begin{adjustbox}{width=0.5\linewidth,center=0pt}
      \includegraphics[width=\linewidth]{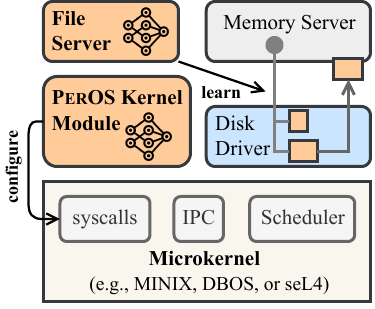}
    \end{adjustbox}
    \caption{Adaptive kernel and subsystems of \sysname{}.}
    \label{fig:kernel}
\end{figure}

%% file: sections/4_deploy.tex
\section{Secure and Scalable Architecture in the Cloud} \label{sec:deploy}

The potential of \sysname{}'s adoption can be significantly improved with a secure and scalable design.
In this post-Moore's Law era, the performance of individual chips is quickly leveling off, challenging the scalability of client-side devices for LLMs and other ML models.
Thus, tapping into the mature cloud ecosystem emerges as a promising solution.
My envisaged approach involves leveraging three key technologies: thin-client computing (TCC), serverless computing, and privacy-preserving ML (PPML).

TCC, proposed over two decades ago, aimed to streamline hardware maintenance and cut costs for deploying and updating applications across diverse users. 
It relies on stateless (thin) clients interacting with centralized computing resources. 
While commercial products like Citrix Metaframe and Microsoft Terminal Services emerged, TCC struggled beyond individual organizations due to strict network requirements and privacy worries from its centralized setup.
Yet, TCC's relevance is resurging due to several key factors. 
Network connectivity is rapidly improving, with speeds doubling roughly every two years since 1983. 
Global 4G coverage is projected to hit nearly 100\% in the next five years, already supporting graphics-heavy mobile apps like 4K@30fps. 
These trends create an ideal environment for TCC's development.
Moreover, the mounting concern over electronic waste highlights the need for sustainable computing. 
TCC stands out by \textit{reducing the need for frequent client device upgrades}, lessening environmental impact from manufacturing these devices and their accessories.
Furthermore, TCC has the potential to extend the \textit{operational duration of client devices}.
As a counterexample, despite battery capacity improvements, the batteries drain much faster in newer generations of iPhones. 
This phenomenon can be attributed partly to the increasingly compute-intensive software and the higher power consumption of on-device ML models.
Additionally, as Moore's Law nears its limits and conventional methods struggle to meet escalating computational demands — especially for emerging AI/ML applications — the cloud's maturation offers a solution. 
Offloading resource-intensive computations to the cloud tackles device limitations, ensuring a sustainable path to handle evolving technology demands.

Serverless computing represents the ongoing evolution in computing, mirroring the trajectory of cost-effective solutions like TCC. 
Typical aerverless services include Function-as-a-Service (FaaS), Database-as-a-Service (DaaS), ML-as-a-Service (MLaaS), etc.
They are built upon the concept of efficient resource utilization and centralized processing, akin to TCC's approach of leveraging centralized computing resources for user interactions. 
In its essence, serverless epitomizes a paradigm shift towards on-demand computing, offering a scalable and resource-efficient model for executing code without the need to manage underlying infrastructure. 
Just as TCC sought to simplify hardware maintenance and reduce deployment costs, serverless continues this trajectory by enabling users to execute individual functions without concerns about the backend infrastructure --- a concept also aligning with the principles of TCC.
While TCC emphasizes thin client interactions with centralized resources, serverless computing prioritizes code execution in a scalable, pay-as-you-go cloud environment. This evolution from hardware-centric to code-centric approaches underscores the ongoing quest for efficiency, cost reduction, and scalability in computing paradigms.

Last but not least, the awareness and plea for data privacy and security are growing. At first glance, this trend is seemingly incongruent with TCC and serverless, wherein computation is centralized and predominantly occurs off-device. 
Nevertheless, recent developments in PPML present a feasible solution for auditing and managing personal data for ML-integrated systems like \sysname{}.
PPML stands at the forefront of ML serving, addressing the critical concern of data privacy and security. 
In contexts like serverless and TCC, where data is centralized or processed in shared environments, preserving the confidentiality of sensitive information becomes paramount. 
PPML employs techniques like cryptography and data auditing to enable collaborative learning without exposing raw data, ensuring that models can be trained and refined without compromising users' privacy. 
Therefore, in serverless and TCC frameworks, PPML could emerge as a crucial safeguard, allowing the development of intelligent systems like \sysname{} while safeguarding the sensitive information integral to these paradigms.

\subsection{Objectives} \label{subsec:obj3}

The objectives here are twofold: \textit{(i)} protecting security and privacy of personal data; \textit{(ii)} ensuring system scalability and facilitating resource sharing.
To enhance the security and privacy of users' personal data, I propose to physically store this information off-cloud in close proximity to the users, a concept inspired by Databox~\cite{yadav2018databox, perera2017databox3, crabtree2018databox2}. 
This approach helps streamline data logging and auditing, pivotal for leveraging PPML techniques.

As most of the computation happens remotely (\S\ref{subsec:method1}), the client-side devices should embody a ``thin" design, relegating most computational tasks to the cloud. 
These devices could range from simple displays to projected screens or envisioned VR/AR headsets in the future. 
Such a TCC-like architecture liberates \sysname{} from concerns about limited compute and storage when utilizing LLMs. 
In contrast, on-device deployment necessitates heavy quantization of ML models (e.g., \cite{Dettmers2023QLoRAEF}), leading to considerable performance degradation.

Importantly, personal devices like mobile phones and laptops \textit{often remain idle} in our pockets or backpacks. 
Hence, a pay-as-you-go billing model, enabled by serverless computing, is crucial for \textit{cost-effectiveness}. 
Additionally, serverless architecture enables users to share common resources. 
For instance, multiple family members can use the same LLM, the Interpreter and Director, as a shared interface service to their \sysname{}es, leveraging the LLM's ability to differentiate users via provided usernames in its prior context. 
Similarly, organizations could share the same LM Manager, aligning retraining policies.
Adopting a serverless approach for hosting \sysname{}es also empowers cloud providers to consolidate server resources and independently scale individual components based on demand fluctuation and for load balancing. 
This strategy can also improve hardware utilization and, in turn, energy efficiency in the cloud.

\subsection{Research Question and Challenges} 
The two objectives elaborated above entail the last RQ of this proposal:
\begin{enumerate} [label=\textit{RQ3}]
 \item \label{rq3} How to enable personalized, ML-powered OSes in the cloud, ensuring scalability to accommodate thousands of users while protecting their privacy and personal data?
\end{enumerate}

\mypar{Challenges}
The challenges induced by \ref{rq3} encompass three main aspects: \textit{(i)} the robustness of the data privacy and security model, \textit{(ii)} concerns regarding performance and latency, and \textit{(iii)} the engineering complexity and deployment costs.

Firstly, while PPML tries to uphold data privacy, centralizing data storage and processing in the cloud under TCC raises apprehensions about potential data breaches or unauthorized access. 
Safeguarding sensitive user information is imperative, needing robust encryption and stringent access control measures. 
These policies aim to mitigate risks such as impersonation and man-in-the-middle attacks. 
Additionally, compliance with data protection regulations like GDPR and HIPAA is important when managing personal data within a cloud-based AI operating system. 
Adhering to these regulations, especially across different jurisdictions, can pose intricate legal and compliance challenges.

Secondly, concerning performance and latency, TCC heavily relies on uninterrupted network connectivity for user interactions with centralized resources. 
Despite substantial improvements in connectivity --- improved, for instance, by offerings like Starlink~\cite{starlink} --- network disruptions or latency issues can still profoundly impact user experience and degrade the responsiveness of the OS. 
Furthermore, offloading computational tasks to the cloud can introduce large latency, especially affecting real-time applications like VR/AR, leading to delays due to the round-trip communication between thin clients and cloud servers.

Thirdly, integrating the triad of technologies --- TCC, PPML, and serverless architecture --- demands overall compatibility and interoperability. 
Ensuring these systems work harmoniously without conflicts presents a significant engineering challenge. 
Moreover, scaling \sysname{} in the cloud via serverless architectures necessitates careful resource orchestration, ensuring adequacy to cater to demand fluctuation while optimizing costs. 
Unoptimized resource allocation might result in unforeseen expenses, as \textit{the net cost of serverless models tends to be much higher} than alternatives such as renting virtual machines (VMs). 
Therefore, cost optimization and resource orchestration remain a critical concern in such deployments.

\subsection{Research Methods} \label{subsec:method3}

\input{assets/fig_deploy}

\mypar{Scope}
To ensure the feasibility of addressing \ref{rq3} within a year, I restrict the research scope as follows.
Firstly, similar to addressing \ref{rq1}, the focus should initially rest on text-only thin-client devices and target applications.
Secondly, although some kernel subsystems involve little state management and can theoretically be shared among trusted tenants, this work should mainly focus on sharing the microkernel (or not sharing any components at the kernel level).
Thirdly, the targeted attack model for this study should remain minimal, explicitly excluding sophisticated attacks such as meltdown and side-channel attacks.
Additionally, this work should operate under the presumption of ample network bandwidth. 
It anticipates a future with significantly improved network connectivity, ensuring suitable conditions both between the client and Databoxes, and between the client and the cloud.
Lastly, specific performance criteria outlined in \S\ref{subsec:method1} and \S\ref{subsec:method2} could be relaxed, if necessary. 
Factors such as adaptiveness and latency, which were initially stringent, might require adjustment due to the increased complexity of the system. 
This adjustment aims to maintain a reasonable balance given the expanded system design.

\mypar{Approach}
To address security and privacy concerns, I plan to first formally define the \textit{threat model} for the proposed system (Figure~\ref{fig:deploy}). 
Specifically, I want to formulate the threat model
from the following three angles.
\textit{(i)} Considering the attacker's viewpoint, threats may originate internally or externally inside or outside the Deployment Groups (e.g., families, organizations).
\textit{(ii)} From an architectural angle, potential vulnerabilities may reside in each stateful system component and the interactions among them. 
Assessing these components involves ranking them by their work factor, i.e., the effort an attacker needs to compromise a component.
\textit{(iii)} Measuring threats from an asset perspective involves considering attack motivations. 
Here, emphasis should be placed on areas housing the most valuable data.
With the threat model defined, the subsequent step involves identifying corresponding indicators of attack (IOAs) and indicators of compromise (IOCs) in the later stages of the research. 
IOAs aid proactive vulnerability prevention, while IOCs shield the system from known security breaches.
Leveraging the computing models embedded in the system design also offers mitigation strategies against potential threats. 
For instance, access to personal data, as per \cite{yadav2018databox}, necessitates explicit examination, authorization, and logging. 
These access traces (\circlew{1}) facilitate data auditing and can be used to automatically sign client-provider contracts in PPML. 
Similarly, deployed as stateless FaaS functions (\circlew{2}, \circlew{6}, \circlew{4}), these system components are less susceptible compared to stateful services.

Subsequently, I intend to combine and extend the prototypes from \S\ref{subsec:method1} and \S\ref{subsec:method2}. 
To align the system with the serverless model, the Actuator, Watchdog (\circlew{3}), and LM Manager (\circlew{4}) should be transformed into standalone functional components to be hosted as FaaS functions (\circley{\textsf{\textbf{fn}}}). 
The Interpreter and Director (\circlew{5}), as shared components, should ideally follow the MLaaS model, although its feasibility necessitates further investigation.
Regarding the microkernel (\circlew{6}), I plan to host it as a normal VM instance due to security and availability concerns. 
Ensuring scalability to meet fluctuating demands (\S\ref{subsec:obj3}), FaaS components should ideally be independently scalable in this design.
Subsystems of \sysname{} that heavily interact with protected user data, like file scanning and storage, should not be shared to uphold data protection in the initial development.

\subsection{Evaluation} \label{subsec:eval3}

Evaluating the proposed system (Figure~\ref{fig:deploy}) includes various factors, including security, performance, scalability, and functionality. 
As the system design is unique, the \textit{baseline} could mirror the proposed design but without any ML models and data protection via Databox, functioning as regular VM clusters.
However, this baseline system will not be able to support declarative requests from users, limiting the comparisons to normal requests like individual system commands and file operations.

Metrics outlined in previous sections (\S\ref{subsec:eval1} and \S\ref{subsec:eval2}) for assessing ML models in the system (such as latency, response time, correctness, and adaptiveness) remain applicable to experiments for addressing \ref{rq3}. 
Yet, these learning metrics might experience degradation due to increased complexity and orchestration overheads (\S\ref{subsec:obj3}).

Adjusting earlier performance experiments to a serverless or cloud context is essential. 
I have extensive experience in working with benchmarks such as DeathStarBench~\cite{gan2019deathstar}, ServerlessBench~\cite{socc20-serverlessbench}, and FunctionBench~\cite{kim2019functionbench}. 
I intend to tailor them for evaluating of \sysname{} for \ref{rq3}. 
These benchmarks will help load-test the system, assessing scalability limits, system responses under stress, and degradation thresholds. 
I want to observe system behavior and auto-scaling mechanisms, evaluating resource utilization levels, scaling times, and resource allocation efficiency.

To assess the threat model, I plan on conducting threat simulations using open-source vulnerability scanning tools like OpenVAS~\cite{openvas} and Nessus~\cite{nessus}, as well as penetration testing tools such as Metasploit~\cite{metasploit} and Nmap~\cite{nmap}. 
In these experiments, I will record successful penetration attempts, access control violations, and identified threats, enabling an extensive comparison with the baseline system.

Finally, comparing the cost of hosting the proposed system over a defined period against the baseline is also informative. 
Utilizing a consistent cost model of offerings from platforms like AWS or Azure will facilitate this assessment. 
Ideally, these experiments should provide comprehensive insights into the system's security, performance, compliance with benchmarks, vulnerability to threats, and cost-effectiveness compared to the baseline setup.

%% file: assets/fig_deploy.tex
\begin{figure}[t]
    \centering
    \begin{adjustbox}{width=1.1\linewidth,center=0pt}
      \includegraphics[width=\linewidth]{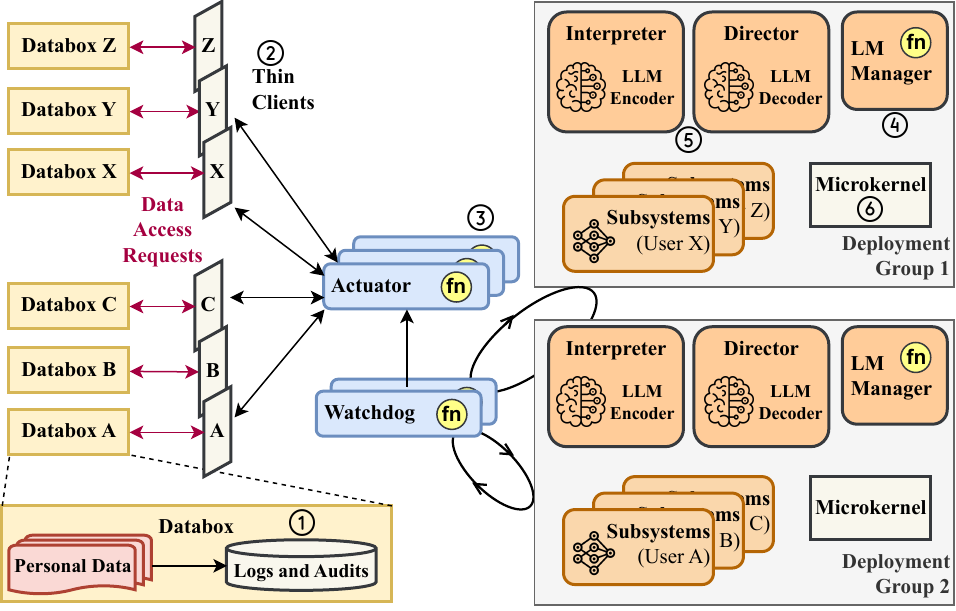}
    \end{adjustbox}
    \caption{\sysname{} architecture overview.}
    \label{fig:deploy}
\end{figure}

%% file: sections/5_related_work.tex
\section{Related Work} \label{sec:related}

\mypar{Declarative system interfaces}
\citeauthor{Etzioni1993agentos} introduced Agent OS~\cite{Etzioni1993agentos}, presenting a goal-oriented approach to OS command interface. 
Unlike traditional step-by-step commands, this model allows users to specify goals, empowering the OS agent to determine command sequences based on system state and its knowledge base. 
It executes user requests through planning techniques. 
Implemented in a distributed UNIX environment, this approach showcases the practical integration of automatic planning and learning into OSes with negligible performance impact.
Another important work is the Internet Softbot~\cite{Etzioni1994softbot}, an AI agent operating with a UNIX shell and the web. 
It interacts with diverse internet resources using commands such as \texttt{ftp}, \texttt{telnet}, \texttt{mail}, and sensor utilities like \texttt{archie}, \texttt{gopher}, and \texttt{netfind}. 
The Softbot dynamically selects and sequences facilities, adapting its behavior based on collected runtime information. 
This adaptation provides an integrated, flexible interface to the internet, capable of responding to user requests like ``Send the budget memos to Mitchell at CMU.''
This paper emphasizes the conceptual foundation of declarative interface.
In contrast, ibot~\cite{Amant2000ibot} takes a different approach, focusing on agents that control interactive applications \textit{directly} through GUIs, rather than application APIs. 
It utilizes a substrate equipped with sensors processing visual data from the display to identify interface elements. 
The interface then generates mouse/keyboard actions to manipulate these elements.
Moving forward, \citeauthor{Jennings1999fish} introduced Fish~\cite{Jennings1999fish}, a command shell fostering intelligent and interactive text-based interfaces. 
It maintains a persistent knowledge repository across concurrent sessions, allowing users to define functions and access previous results. 
Similarly, \citeauthor{Copas2000IntelligentIT} described interactive interfaces~\cite{Copas2000IntelligentIT} in planning systems, such as geographic information systems, known for their high functionality but poor usability.
Furthermore, \citeauthor{Petrick2007PlanningFD} explored using DCOP (Desktop COmmunication Protocol) to link desktop application services, employing knowledge-level conditional planners to control KDE applications~\cite{Petrick2007PlanningFD}. 
This research demonstrates executable plans for application manipulation and information retrieval, underscoring the potential of desktop interfaces for automated planning in OSes.

Unfortunately, these declarative interfaces and softbots from previous decades heavily rely on preset rules, primarily subsets of first-order logic stored in a knowledge base~\cite{Kruit2020Tab2KnowBA} containing operation descriptions and related semantics. 
The ``AI'' or ``learning'' described in these papers revolves around planning based on reasoning and derivation of logical expressions. 
Consequently, these interfaces lack adaptivity to shifts in user usage patterns, extensibility to new APIs and functionalities, as well as the smooth multi-turn conversational capability of LLMs (\S\ref{subsec:obj1}).


\smallskip
\mypar{ML for system design and modeling}
There is a large body of work employing ML models for system design and modeling, ranging from low-level hardware design (e.g., circuit analysis~\cite{Pan2019LateBR,Alawieh2020HighDefinitionRC}, network-on-chip~\cite{Reza2018NeuroNoCEO,Boyan1993PacketRI,Joardar2018LearningBasedA3}, and logic synthesis~\cite{Liu2013OnLM,Dai2018FastAA,Ustun2020AccurateOD}) to high-level resource allocation and management (e.g., kernel scheduling~\cite{Whiteson2005CORRECTEDV,Fedorova2007OperatingSS,Vengerov2009ARL}, power management~\cite{Reza2018NeuroNoCEO,Fettes2019DynamicVA,Dinakarrao2016AQB,Juan2012PowerawarePI,Imes2018EnergyefficientAR}, and cloud orchestration~\cite{Zhang2018LearningDP,Zhang2021SinanMA,Zhou2022AQUATOPEQR,Wang2022SOLSO}).
Here, I elaborate more on existing work related to memory and storage systems for \ref{rq2}

In the realm of memory systems, exploiting ML-based performance models proves instrumental in examining the tradeoffs across various objectives. 
\citet{Dong2013ACC} delve into NVM-based cache hierarchies, using an ML model to predict higher-level features like cache read/write misses and instruction-per-cycle rates from lower-level attributes such as cache associativity, capacity, and latency. 
Block2Vec~\cite{Dai2018FastAA} trains an ML model to derive optimal vector representations for individual data blocks, thereby uncovering block correlations and enabling enhanced caching and prefetching optimizations through vector distances. 
In a similar vein, \citet{Shi2019LearningET} harness a graph neural network to mix static code and dynamic execution into a unified representation. 
This approach models both data flows, such as prefetching, and control flows, including branch prediction, offering a comprehensive understanding of program behavior.
A recent work that caught my eye was LeaFLT~\cite{sun2023leaftl}, a learning-based flash translation layer for SSDs. 
LeaFTL employs runtime linear regression to dynamically learn address mappings, reducing memory usage for addressing tables, enhancing data caching in SSD controllers and demonstrating speedup in storage performance compared to existing flash translation schemes, while consuming less memory for mapping tables.

Although the abundant literature on this topic offers valuable insights into applying ML methods to systems research, their primary focus remains on enhancing the performance of individual system components rather than facilitating personalized user experiences at the OS level. 
In contrast, \ref{rq2} emphasizes the collaborative synergy among kernel components, spanning from high-level filesystem configurations and allocation policies down to learning to index data blocks of a file. 
The proposed approach aims to adapt proactively to individual usage patterns, thereby tailoring the OS for personalized experience.
Furthermore, there exists a notable gap in the exploration of ML methodologies to enhance filesystem operations holistically. 
The earliest reference I found, dating back decades, is SUMPY~\cite{Song2007SUMPYAF}, a software agent for UNIX filesystem maintenance. 
SUMPY focuses on optimizing disk space through file compression and backup. 
Its layered structure enables concurrent goal pursuit, resilience against failures, and the addition of new tasks. 
However, it relies on basic fuzzy logic, which restricts its extensibility and limits its scope to automating rudimentary background tasks.

\smallskip
\mypar{TCC, Serverless, and PPML}
In recent years, research in TCC has been relatively scarce, primarily focusing on performance evaluations of existing solutions (e.g., \cite{Schlosser2010ImprovingTQ, Casas2013QualityOE, nieh2000comparison, maga2013comparison, yang2002performance, schmidt1999interactive}) and specialized systems tailored for specific use cases (e.g., \cite{Magaa2019RemoteAP, deboosere2007thin, suznjevic2015statistical, Ali2009ReducingPC, Simoens2011OptimizedMT, Lee2006ThinClientCF}). 

In contrast, serverless computing has gained significant attention, exploring various aspects such as minimizing cold start overhead~\cite{Silva2020PrebakingFT, Shahrad2020ServerlessIT, Shin2022FireworksAF, Ustiugov2021BenchmarkingAA, Roy2022IceBreakerWS, Agache2020FirecrackerLV}, function scheduling~\cite{Mahmoudi2022PerformanceMO, Kaffes2022HermodPA, Tang2022DistributedTS, Aslanpour2022EnergyAwareRS, Bhasi2022CypressIS,invitro}, and workflow orchestration~\cite{Mahgoub2022ORIONAT, Pons2022StatefulSC, Mahgoub2022WISEFUSEWC, Burckhardt2022NetheriteEE, Risco2021ServerlessWF}. However, there is an absence of attention towards providing personalized, ML-enhenced OSes in the cloud as stateful services. 
Moreover, no prior work has attempted to scale similar services by sharing stateless components of the hosted stateful applications.

Concerning data protection for internet-accessible applications, practical systems like the Databox~\cite{yadav2018databox, perera2017databox3, crabtree2018databox2} and theoretical models like the Databank~\cite{hublet2021databank} offer promising solutions. 
Databox includes physical and cloud-based software components designed to empower individuals in managing, logging, and monitoring access to their personal data by third parties. 
Similarly, the Databank model showcases formal guarantees on information flow propagation and policy enforcement, offering insights into monitoring algorithms and prototype infrastructure.

Within the sphere of ML, PPML has emerged as a rapidly evolving field, encompassing techniques like cryptographic hashing, secure computation for inference, and non-colluding ML servers. 
For instance, platforms like Piranha~\cite{Watson2022PiranhaAG} accelerate secret sharing-based multi-party computation protocols using GPUs. 
Meanwhile, MUSE~\cite{Lehmkuhl2021MUSESI} introduces a resilient two-party secure inference protocol employing novel cryptographic approaches. 
SecureML~\cite{Mohassel2017SecureMLAS} innovates protocols for PPML in various domains like linear and logistic regression, as well as neural network training, ensuring privacy while jointly training models across non-colluding servers.

The system model of Databox offers valuable insights into data access logging and auditing, which could be leveraged in theoretical models such as the Databank. 
I believe that a low-hanging fruit in the PPML domain lies in efficiently signing contracts between data holders (e.g., users) and model holders (e.g., cloud providers) based on detailed logs and audits for the requests and accesses of users' personal data.